\documentclass[aps,prb,twocolumn,superscriptaddress]{revtex4}
\usepackage{ae}
\usepackage[T1]{fontenc}
\usepackage[ansinew]{inputenc}
\usepackage{amsmath}
\usepackage{amssymb}
\usepackage{graphicx}
\usepackage{color}
\usepackage[colorlinks]{hyperref}
\usepackage{epstopdf}

\begin{document}

\title{Peierls-type structural phase transition in a crystal induced by
 magnetic breakdown}
\author{A. M. Kadigrobov}\affiliation{Department of Physics, Faculty of Science,
University of Zagreb, POB 331, 10001 Zagreb,
Croatia}\affiliation{Theoretische Physik III, Ruhr-Universit\"{a}t
Bochum, D-44801 Bochum, Germany}
\author{A. Bjeli\v{s}}\affiliation{Department of Physics, Faculty of Science, University of
Zagreb, POB 331, 10001 Zagreb, Croatia}
\author{D. Radi\'{c}}\affiliation{Department of Physics, Faculty of Science, University of
Zagreb, POB 331, 10001 Zagreb, Croatia}

%\date{\today}

\begin{abstract}
We predict a new type of phase transition in a
quasi-two dimensional system of  electrons at high magnetic fields, namely
the stabilization of a density wave which transforms a two
dimensional open Fermi surface into a periodic chain of large
pockets with small distances between them. The quantum tunneling of electrons between the
neighboring closed orbits enveloping these pockets transforms the
electron spectrum into a set of extremely narrow energy bands and
gaps which decreases the total electron energy, thus leading to a
\emph{magnetic breakdown induced density wave} (MBIDW) ground state. We show that this DW instability has some qualitatively different properties
in comparison to analogous DW instabilities of Peierls type. E. g. the critical temperature
of the MBIDW phase transition arises and disappears in a peculiar way with a change of the
inverse magnetic field.
\end{abstract}

\maketitle

\section{Introduction}

A specific topology of an open Fermi surface with two corrugated sheets related by the reflection symmetry is the origin of
numerous examples of modulated ground states with generally incommensurate periodicities. These are well known charge or spin
density wave (DW)  ground states, widely investigated during last decades. DW ground states are observed in materials with a quasi-one-dimensional crystal lattice, or materials which, due to other reasons, show strong internal anisotropy of conducting band
electronic states. \cite{reviewDW}

In the absence of external magnetic field the longitudinal component $Q_{x}$ of the DW momentum $\textbf{Q}$ has a value that is
equal or close to $2p_{F}$, with Fermi momentum $p_{F} = n \hbar \pi/a$ where $a$ is the longitudinal lattice constant, and $n$ is the number of electrons
per site filling the band for which only the spin degeneracy is assumed. As is seen in Fig. \ref{trajectories}, $2p_{F}$ is the mean
distance between two ("upper" and "lower") Fermi sheets.

The DW in the absence of magnetic field may be stabilized provided one geometric and one energetic
condition are obeyed. Geometrically, its transverse component $Q_{y}$ has to respect the nesting requirement, i. e. to allow for
the largest possible phase space for the condensation of electron-hole pairs carrying the momenta $\bf{\pm Q}$. Correspondingly,
the great part of the new Fermi surface is gapped. Still, the nesting in real materials is imperfect, so that the Fermi surface of
DW state contains mutually distant pockets in which the DW gap is degraded or even vanishes. Energetically, the electron-electron or
electron-phonon coupling responsible for the condensation of electron-hole pairs, have to be strong enough in order to get
a sufficient correlation energy gain and stabilize the DW order.

Even if the above conditions are not obeyed, field-induced density wave (FIDW) orderings may be stabilized after applying a
strong enough external magnetic field $H$ perpendicular to the $(x,y)$ plane. As was pointed out in our previous work \cite{ourwork}, two
possibly competitive mechanisms may lead to FIDWs.

The first one may be realized if the nesting is not far from perfect, i. e. in the regime $t_{y}' \ll t_{y}$,
where $t_{y}$ is the transverse band width, and $t_{y}'$ is the imperfect nesting parameter. It is based on the
"one-dimensionalization" of electron spectrum due to the Landau quantization of band states within distant, presumably
small, pockets remaining after establishing the $2p_{F}$ modulation.\cite{GorLeb,reviewFISDW}
The magnetic field energy relevant for this type of FIDWs is characterized by the scale of cyclotron energy which is in the present case given by $\hbar \omega_{c}$ with $\omega_{c}= e v_{F}H/b^{*}c$ where $v_{F}$ is the longitudinal Fermi velocity, and $b^{*}$ is the transverse reciprocal lattice constant in momentum space. FIDWs then can be stabilized provided $\hbar \omega_{c}$ is of the order of
imperfect nesting parameter $t_{y}'$.
This requirement is accessible with magnetic fields of the order of up to few tens tesla for various families of quasi-one-dimensional and quasi-two dimensional materials like Bechgaard salts, $\alpha$-ET compounds, etc.

While in this regime the effects of tunneling between distant pockets are negligible, the alternative choice of DW period, for which the pockets are large and barriers between them narrow, leads to a novel mechanism of DW stabilization. It originates from the decrease of the total electron energy due to the creation of finite barriers (instead of simple crossing points in the absence of DW) between neighboring pockets. The energy decrease comes from the magnetic breakdown between neighboring pockets which opens the gaps in one-dimensional sub-bands, the latter appearing due to the orbital quantization in the case of an open (quasi-one-dimensional) conducting band. Such finite barriers go together with the formation of lattice periodic modulation, which in turn counterbalances the band energy gain with the increase of the lattice energy. The outcome is the ordered DW, stabilized by the magnetic field assisted tunneling, i. e. by the one-particle processes, and not by the electron-electron scattering like in the regime of nested DWs. On the contrary, the electron-electron correlations, illustrated here by the electron-phonon coupling, oppose, and eventually equilibrate, the stabilization of MBIDWs.

With magnetic breakdown having the central role in this kind of DW ordering, it is necessary to investigate how the details characterizing the barriers influence the electron spectrum and the free energy of the ordered state. More precisely, one comes to the problem of finding the barrier configuration that is the most favorable one for the DW stabilization, leading to the highest corresponding critical temperature.

In the previous work \cite{ourwork}, we have considered the case of the DW momentum equal to $(2p_F, 0)$, i. e. the succession of same pockets separated by a simple point-like barriers without internal structure, the distance between neighboring barriers being equal to $b^{*}/2$. In the present analysis we concentrate to the ordering with the DW momentum roughly equal to $(2p_F - 4t_y/v_F,0)$. This is the regime of largest possible pockets, i. e. the limit of full "anti-nesting" shown in Fig. \ref{trajectories} The distance between neighboring barriers is now equal to $b^{*}$.
In the latter choice the "upper" and "lower" sub-bands do not cross but touch each other, so that the large pockets are not separated by barriers, but by small pockets with a peculiar local band dispersion. As our present analysis shows, these small pockets play the role of effective magnetic barriers with a qualitatively enhanced effect of field assisted tunneling between large pockets. Consequently, we come to the central result of this work, namely that the fully anti-nesting regime from Fig. \ref{trajectories} is the best candidate for the MBIDWs.\\

The overview of the paper is as follows: In Chapter II we present the spectrum and the density of states of quasi-one-dimensional electron band with periodic perturbation introduced by the DW formation, all under magnetic field with magnetic breakdown (MB) induced tunneling between electron trajectories treated within the framework of semiclassical formalism. In Chapter III we calculate the energy balance between energy loss of delocalized electron due to the MB tunneling and gain due to the DW formation that leads to the magnetic breakdown induced transition of Peierls type. There we predict the magnetic breakdown induced density wave (MBIDW) and present its phase diagram on the domain of magnetic field and critical temperature. Concluding remarks are given in Chapter IV. An Appendix, provided in the end, contains mathematical details related to the calculation of results from the main text, namely App. \ref{dynamics} for Chapter II and App. \ref{fourierapp} for Chapter III. In Section \ref{dynamics}-1 we present the calculation of the resulting band structure of quasi-one-dimensional band under the periodic perturbation. Section \ref{dynamics}-2 contains the details of quantum-mechanical calculation of novel MB tunneling process for the particular bands relevant for the present problem. Furthermore, in Section \ref{dynamics}-3 we present the semiclassical formalism used to describe electrons apart from the MB region and calculate the new band structure under the regime of magnetic breakdown. Finally, in \ref{dynamics}-4 we show the construction of the MB tunneling matrix used to connect the semiclassical solutions from the different regions and provide the novel tunneling probability for the proposed model. Appendix \ref{fourierapp} contains the details of the Fourier expansion of electron density of states under the MB regime.

\section{Spectrum and density of states of electrons under magnetic field}

We consider a metal with an open Fermi surface assuming that the dispersion law
of conduction electrons  $\varepsilon (p_x, p_y)$ depends only
on two projections of the electron quasi-momentum $\textbf{p}=(p_x,
p_y)$. It can be the case either of a three dimensional metal, with
a weak dependence of its dispersion law  on $p_z$ as it is in
two-dimensional conductors, or highly anisotropic quasi-one-dimensional conductors like Berchgaard salts.
The open Fermi surface $\varepsilon (p_x, p_y)= \varepsilon_F$ consists of two branches, "$+$" (upper) and "$-$" (lower),
\begin{equation}
\varepsilon_{\pm}(\textbf{p}) = \varepsilon_F \pm v_F (p_x \mp p_F) - 2t_y cos\frac{b p_y}{\hbar},
\label{Bechgaard_dispersion}
\end{equation}
where $\varepsilon_F$ and $v_F$ are the Fermi energy and velocity respectively (see Appendix \ref{dynamics}, Eqs. (\ref{initialBechgaard}) - (\ref{energies}) for details).\\
%
%%%%%%%%%%%%%%%%%%%%%%%%%%%%%%%%%%%%%%%%%%%%%%%%%%%%%%%%%%%%%%%%%%%%%%%%%%%%%
\begin{figure}
\centerline{\includegraphics[width=0.95\columnwidth]{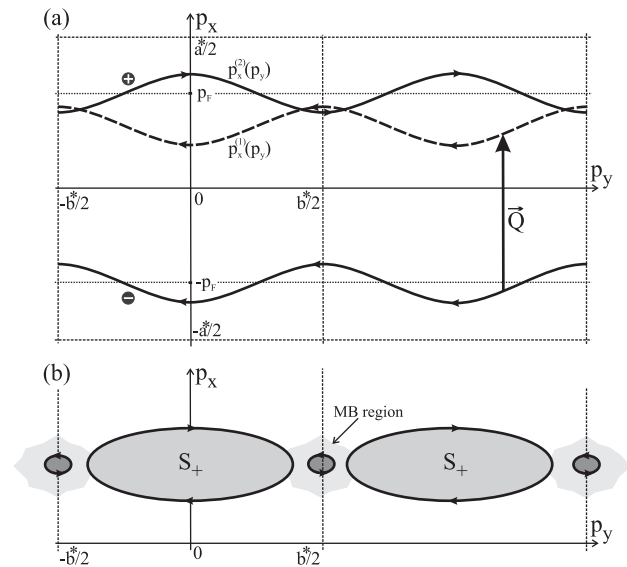}}
\caption{(a) An open electron Fermi surface, with two branches "$+$" and
"$-$", shifted by deformation momentum $\textbf{Q}$ into $p_x^{(1,2)}(p_y)$. Arrows depict electron trajectories with the
opposite directions of motion at corresponding branches in magnetic
field. Expressions that determine the shape of trajectories $p_x^{\pm}(p_y)$ are
given by Eq. (\ref{iso-energetic-sheets}) in Appendix. $a^*\equiv 2\pi\hbar / a$ and $b^* \equiv 2\pi\hbar / b$ are the reciprocal lattice
constants in $x$- and $y$-direction of momentum space, respectively, corresponding to the real space lattice constants $a$ and $b$. (b) After the degeneracy is lifted at the
crossing points due to DW potential, the chain with two types of
electron orbits is formed: the large ones (with area $S_+$ in
p-space) and the small ones between them. In the finite magnetic
field, the magnetic breakdown between neighboring $S_+$ trajectories
takes place through the area occupied by small pockets (MB-regions).}
\label{trajectories}
\end{figure}
%%%%%%%%%%%%%%%%%%%%%%%%%%%%%%%%%%%%%%%%%%%%%%%%%%%%%%%%%%%%%%%%%%%%%%%%%%%%%
%
We assume that a static periodic lattice distortion creates a
modulation potential $V(x)$ with period $2\pi \hbar /Q_x$. The
deformation momentum $\textbf{Q} \approx (2p_F - 4t_y/v_F,0)$, with the optimal value to be determined later from the DW condensation condition, combines open trajectories
$p_x^{(+)}(p_y)$ and $p_x^{(-)}(p_y)$ with opposite directions of
the electron motion. These trajectories are the upper and the lower branches of
the Fermi surface depicted by
thick solid lines in Fig. \ref{trajectories}(a). Therefore, the combined trajectories  are
%
%%%%%%%%%%%%%%%%%%%%%%%%%%%%%%%%%%
\begin{eqnarray}
p_x^{(1)}(p_y)&=&p_x^{(-)}(p_y) +Q_x \nonumber \\
p_x^{(2)}(p_y)&=&p_x^{(+)}(p_y).
\label{pshift}
\end{eqnarray}
%%%%%%%%%%%%%%%%%%%%%%%%%%%%%%%%%%
%
We show below that the most favorable MBIDW deformation
vector $Q_x$ is close to the one at  which the  trajectories
 $p_x^{(1)}(p_y)$ and $p_x^{(2)}(p_y)$ touch  each other. Hence,
the new energy spectrum $E_{1,2} (p_x,p_y)$ of the crystal in the presence of the modulation potential $V(x)$ has peculiar points in the energy
space near the Fermi energy $\varepsilon_F$ at which the equipotential surfaces $E_{1,2}(p_x,p_y)=\varepsilon$ change their topology with a change of the electron energy $\varepsilon$.
The details of electron spectrum in the vicinity of these points are presented in Appendix \ref{dynamics}, Eq. (\ref{newenergies}), Fig. \ref{configurations1}.

Here we consider dynamics of electrons in the presence of the modulation potential $V(x)$ under a strong magnetic field $H$
at which the Larmor radius is much smaller than the electron free
path length $l_0$, that is
%
%%%%%%%%%%%%%%%%%%%%%%%%%%%%%%%%%%
\begin{eqnarray}
\frac{c p_F}{e H} \ll l_0,
\label{Larmor}
\end{eqnarray}
%%%%%%%%%%%%%%%%%%%%%%%%%%%%%%%%%%
%
where $c$ is the light velocity and $e$ is the electron charge. On the other hand, we
assume that the magnetic field satisfies the inequalities
%
%%%%%%%%%%%%%%%%%%%%%%%%%%%%%%%%%%
\begin{eqnarray}
S_{small} < \sigma \ll S_+;\hspace{0.5cm} \sigma \equiv \frac{e\hbar H}{c},
\label{}
\end{eqnarray}
%%%%%%%%%%%%%%%%%%%%%%%%%%%%%%%%%%
%
where $S_{small}$ and $S_+$  are the areas of the small and large
orbits within the trajectory chain respectively (see Fig. \ref{trajectories}(b)). In this
case the electron dynamics may be treated semiclassically between
the MB regions around the small pockets
(marked area in Fig. \ref{trajectories}(b)) in which the
semiclassical approximation is not valid. As we show in Appendix
\ref{dynamics} - section 2-4, Eq. (\ref{rsquared1}), in these regions a peculiar combination of intra-
and inter-band MB transitions results in the
dependence of the MB probability on the magnetic
field and the Fermi energy that qualitatively differs from the
conventional one \cite{Blount} and, as we prove in Chapter \ref{MBPT}, it is the most favorable for the stabilization of the density wave.

In order to find  the wave function of an electron
$G(P_{x0},P_y)$ in the momentum representation between the MB regions, one may use the Onsager-Lifshitz Hamiltonian \cite{Lifshits}
%
%%%%%%%%%%%%%%%%%%%%%%%%%%%%%%%%%%
\begin{eqnarray}
E (P_{x0}+i \sigma \frac{d }{d P_y}, P_y)G(P_{x0},P_y)=\varepsilon G(P_{x0},P_y),
\label{LO}
\end{eqnarray}
%%%%%%%%%%%%%%%%%%%%%%%%%%%%%%%%%%%
%
where  $E(p_x, p_y)$ is the electron energy band in the absence of
the magnetic field, corresponding to the large pocket in Fig.
\ref{trajectories}(b), ${\bold P}$ is the generalized momentum,
$P_{x0}$ is its $x$ component  conserved in the
Landau gauge of the vector potential ${\bold A} = (-H y, 0,0)$.
As the chain of the trajectories under consideration is periodic, the
Hamiltonian (\ref{LO}) is supplemented with a periodic boundary
condition
%
%%%%%%%%%%%%%%%%%%%%%%%%%%%%%%%%%%
\begin{eqnarray}
G(P_{x0},P_y)=G(P_{x0},P_y+b^\ast) \label{Periodic},
\end{eqnarray}
%%%%%%%%%%%%%%%%%%%%%%%%%%%%%%%%%%
%
where  $b^\ast$ is the period of the reciprocal lattice in the
$P_y$-direction. In addition, in the MB regions,
the eigenfunctions are coupled  by the MB condition
presented below.

\begin{figure}
\centerline{\includegraphics[width=0.5\columnwidth]{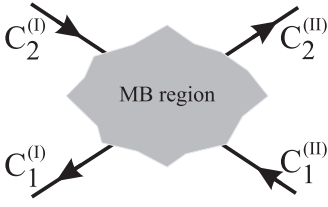}}
\caption{Graphical presentation of the magnetic breakdown scattering proces. The
directions of  arrows show the motion directions of electrons;
$C_2^{(I)}$, $C_1^{(II)}$ and $C_1^{(I)}$, $C_2^{(II)}$ are the
constant factors in the wave functions of the incoming and outgoing
electrons respectively.}
\label{MBmatrix1}
\end{figure}

The semiclassical solution of Eq.(\ref{LO}) can be written in the form
%
%%%%%%%%%%%%%%%%%%%%%%%%%%%%%%%%%%
\begin{eqnarray}
G^{(I)}_{1,2}= \frac{C^{(I)}_{1,2}}{\sqrt{|v_{1,2}|}}
 \exp\left\{-\frac{i}{\sigma}\int_{-\frac{b^\ast}{2}}^{P_y}\left(p_x^{(1,2)}(P_y^\prime)-P_{x0}\right)dP_y^\prime\right\}
\label{WFI}
\end{eqnarray}
%%%%%%%%%%%%%%%%%%%%%%%%%%%%%%%%%%
%
in the region $-3b^\ast /2 < P_y < -b^\ast /2$ (region I),  and
%
%%%%%%%%%%%%%%%%%%%%%%%%%%%%%%%%%%
\begin{eqnarray}
G^{(II)}_{1,2}= \frac{C^{(II)}_{1,2}}{\sqrt{|v_{1,2}|}}
 \exp\left\{-\frac{i}{\sigma}\int_{-\frac{b^\ast}{2}}^{P_y}\left(p_x^{(1,2)}(P_y^\prime)-P_{x0}\right)dP_y^\prime\right\}
 \label{WFII}
\end{eqnarray}
%%%%%%%%%%%%%%%%%%%%%%%%%%%%%%%%%%
%
in the region $ -b^\ast /2 < P_y < b^\ast /2$ (region II), with $v_{1,2} = \partial
E/\partial p_x$ at $p_x=p_x^{(1,2)}(P_y)$. Here and below we treat
the MB regions as point-like objects, neglecting their length in comparison with
the length of the semiclassical trajectories between them. Note that the quantum inter-band transitions between the new energy bands $E_1(p_x,p_y)$ and $E_2(p_x,p_y)$ under the  magnetic field are significant for the values of $p \sim \sqrt{\sigma} \ll b^\ast $ \cite{SK,S}. The
dependence of the kinematic momentum $p_x^{(1,2)}(P_y)$ on $P_y$ is found from the equation
%
%%%%%%%%%%%%%%%%%%%%%%%%%%%%%%%%%%
\begin{eqnarray}
E (p_x, P_y)=\varepsilon,
 \label{epsilon=E}
\end{eqnarray}
%%%%%%%%%%%%%%%%%%%%%%%%%%%%%%%%%%
%
where constants $C^{(I,II)}_{1,2}$ are matched by the $2
\times 2$ MB matrix
%
%%%%%%%%%%%%%%%%%%%%%%%%%%%%%%%%%%
\begin{eqnarray}
\left(
  \begin{array}{c}
    C_1^{(I)} \\
    C_2^{(II)} \\
  \end{array}
\right) =e^{i \Theta} \left(
  \begin{array}{cc}
    t & r\\

  -r^\ast & t^\ast \\
  \end{array}
\right) \left(
  \begin{array}{c}
    C_1^{(II)} \\
    C_2^{(I)} \\
  \end{array}
\right).
\label{MBEquatopns}
\end{eqnarray}
%%%%%%%%%%%%%%%%%%%%%%%%%%%%%%%%%
%
Due to the unitarity of MB matrix, the matrix elements satisfy $|t|^2 + |r|^2=1$, where $|t|^2$ and $|r|^2$ are the MB probabilities for an incident electron to pass through or to be reflected at the
MB region respectively (see Fig. \ref{MBmatrix1}), $\Theta$ is the phase factor specific for the particular MB configuration.  For the case under our consideration, the MB matrix is found in Appendix \ref{dynamics} - section 4, see  Eqs.(\ref{MBmatrix},\ref{rsquared}) .
%
%%%%%%%%%%%%%%%%%%%%%%%%%%%%%%%%%%%%%%%%%%%%%%%%%%%%%%%%%%%%%%%%%%%%%%%%%%%%%
%%%%%%%%%%%%%%%%%%%%%%%%%%%%%%%%%%%%%%%%%%%%%%%%%%%%%%%%%%%%%%%%%%%%%%%%%%%%%
%
%
Using the boundary condition (\ref{Periodic}), one finds
%
%%%%%%%%%%%%%%%%%%%%%%%%%%%%%%%%%%
\begin{eqnarray}
C_1^{(I)}=C_1^{(II)}\exp{\left\{-\frac{i}{\sigma}\int_{-\frac{b^\ast}{2}}^{\frac{b^\ast}{2}}\left(p_x^{(1)}(P_y^\prime)-P_{x0}\right)dP_y^\prime\right\}}\nonumber\\
C_2^{(I)}=C_2^{(II)}\exp{\left\{-\frac{i}{\sigma}\int_{-\frac{b^\ast}{2}}^{\frac{b^\ast}{2}}\left(p_x^{(2)}(P_y^\prime)-P_{x0}\right)dP_y^\prime\right\}}.
\label{PeriodEquations}
\end{eqnarray}
%%%%%%%%%%%%%%%%%%%%%%%%%%%%%%%%%%
%
Equations (\ref{MBEquatopns}) and (\ref{PeriodEquations}) are a set of four
homogeneous algebraic equations  for four unknowns $C_{1,2}^{(I,II)}$ with
the determinant given by
%
%%%%%%%%%%%%%%%%%%%%%%%%%%%%%%%%%%
\begin{eqnarray}
D(\varepsilon, P_{x0})=\cos\frac{ S_+(\varepsilon)}{2\sigma}-|t|\cos\frac{S_-(\varepsilon) -
2b^\ast P_{x0}}{2\sigma},
 \label{D}
\end{eqnarray}
%%%%%%%%%%%%%%%%%%%%%%%%%%%%%%%%%%
%
where
%
%%%%%%%%%%%%%%%%%%%%%%%%%%%%%%%%%%
\begin{equation}
S_{\pm}= S_2 \pm S_1
\label{Spm}
\end{equation}
%%%%%%%%%%%%%%%%%%%%%%%%%%%%%%%%%
%
are the semiclassical actions defined in general as
%
%%%%%%%%%%%%%%%%%%%%%%%%%%%%%%%%%%
\begin{equation}
S_1 = \int_{\frac{b^\ast}{2}}^{-\frac{b^\ast}{2}} p_x^{(1)}(P_y) dP_y, \hspace{0.5cm} S_2= \int_{-\frac{b^\ast}{2}}^{\frac{b^\ast}{2}}
p_x^{(2)}(P_y)dP_y.
\label{actions}
\end{equation}
%%%%%%%%%%%%%%%%%%%%%%%%%%%%%%%%%
%
%
Therefore, the dispersion law of an  electron under magnetic field,
moving along the chain of semiclassical trajectories connected by MB
regions (see  Fig. \ref{trajectories}(b)), is
determined by the electron dispersion equation
%
%%%%%%%%%%%%%%%%%%%%%%%%%%%%%%%%%%
\begin{eqnarray}
D(\varepsilon, P_{x0})=0 \label{D1}.
\end{eqnarray}
%%%%%%%%%%%%%%%%%%%%%%%%%%%%%%%%%%

From here it is obvious that the electron spectrum depends on a discrete
number $n$ and the electron momentum $P_{x0}$. The spectrum
has a band structure $E_n (P_{x0}), \;n=0,1, 2, ...$,   which is
periodic in $P_{x0}$ with the period $\delta P= 2\pi \sigma /b^\ast \ll
b^\ast, a^\ast$. The bands are separated by energy gaps determined by the
condition
%
%%%%%%%%%%%%%%%%%%%%%%%%%%%%%%%%%%
\begin{eqnarray}
\left|\cos{\frac{S_+(\varepsilon)}{2\sigma}}\right| \geq |t|.
\label{gap}
\end{eqnarray}
%%%%%%%%%%%%%%%%%%%%%%%%%%%%%%%%%%
%
In order to find the electron density of states (DOS), we use the
approach developed by Slutskin for spectra of electrons under
magnetic breakdown (see review paper \onlinecite{Slutskin}). Using the identity
%
%%%%%%%%%%%%%%%%%%%%%%%%%%%%%%%%%%%
\begin{equation}
\sum_n \delta(\varepsilon -E_n)=\left| \frac{\partial D}{\partial \varepsilon} \right|\delta(D)
\label{identity}
\end{equation}
%%%%%%%%%%%%%%%%%%%%%%%%%%%%%%%%%%%
%
(see Eqs.(\ref{D},\ref{D1})), the electron DOS
%
%%%%%%%%%%%%%%%%%%%%%%%%%%%%%%%%%%%
\begin{equation}
\nu(\varepsilon)=\frac{1}{L_y}\int_0^{\Delta P_{x}}  \sum_n \delta \left(\varepsilon - E
_n(P_{x0})\right)\frac{d P_{x0}}{2 \pi \hbar }
\label{dos}
\end{equation}
%%%%%%%%%%%%%%%%%%%%%%%%%%%%%%%%%%%
%
can be rewritten in the form
%
%%%%%%%%%%%%%%%%%%%%%%%%%%%%%%%%%%%
\begin{equation}
\nu (\varepsilon) =\frac{1}{L_y}\int_0^{\Delta P_{x}} \left|\frac{\partial
D(\varepsilon, P_{x0})}{\partial \varepsilon}\right| \delta \Bigl(D(\varepsilon, P_{x0})\Bigr)
 \frac{d
P_{x0}}{2 \pi \hbar},
\label{nuD}
\end{equation}
%%%%%%%%%%%%%%%%%%%%%%%%%%%%%%%%%%%
%
where $L_y$ is the width of the sample in $y$-direction and $\Delta
P_x = (eH/c)L_y$ is the maximal value of $P_{x0}$  at which the
center of localization of the electron wave function remains inside
the sample.

Substituting expression (\ref{D}) into (\ref{nuD}) and integrating the latter
with respect to $P_{x0}$, one finds the density of states
%
%%%%%%%%%%%%%%%%%%%%%%%%%%%%%%%%%%%
\begin{equation}
\nu (\varepsilon)= \frac{2 |S_+^\prime|}{(2 \pi \hbar)^2 }
\frac{\left|\sin{\Phi_+(\varepsilon)}\right|}{\sqrt{|t|^{2}-\cos^2{\Phi_+(\varepsilon)}}}
\theta \left(|t|^{2}-\cos^2{\Phi_+(\varepsilon)}  \right),
\label{density}
\end{equation}
%%%%%%%%%%%%%%%%%%%%%%%%%%%%%%%%%%%
%
where $\Phi(\varepsilon)=S_+(\varepsilon)/2\sigma$ and $S_+^\prime = d S_+ /d \varepsilon$. For
example, for the initial spectrum of electrons (\ref{Bechgaard_dispersion}), one has
%
%%%%%%%%%%%%%%%%%%%%%%%%%%%%%%%%%%%
\begin{eqnarray}
S_+ (\varepsilon)= 2b^\ast (\varepsilon - \varepsilon_F+2t_y)/v_F, \;\;S_+^\prime = 2b^\ast/v_F.
\label{S and S prime}
\end{eqnarray}
%%%%%%%%%%%%%%%%%%%%%%%%%%%%%%%%%%%
%
Therefore, there are energy "gaps" in the density of states
determined by Eq.(\ref{gap}). The widths of the gaps are of the order
of $|r|^2 \sigma / S_+^\prime =|r|^2 \hbar \omega_c$, while the width
of the energy bands is of the order of $|t|^2  \hbar \omega_c$, where $\omega_c\equiv
\frac{eH}{m_H^\ast c}$ is the electron cyclotron frequency and $m_H^\ast \equiv S_+^\prime$ plays the role of electron cyclotron effective mass for a semiclassical motion of the electron under an external magnetic field. We show below that such a dramatic transformation of the electron spectrum under magnetic breakdown can result in a peculiar instability of Peierls type.

\section{Magnetic breakdown induced Peierls transition}\label{MBPT}

We have assumed that a static distortion of the crystal combines
open trajectories into a chain of closed orbits with small MB regions between them as it is shown in Fig.\ref{trajectories}.  As a
result,  the initially continuous electron spectrum transforms
into a series of alternating narrow energy gaps  and bands, hence the Fermi energy of the electron system
should inevitably attain a new position on the energy scale. We
find a new value of the Fermi energy $\varepsilon_F$ from the
condition that the number of electrons conserves under the MB induced
Peierls-type phase transition, that is
%
%%%%%%%%%%%%%%%%%%%%%%%%%%%%%%%%%%
\begin{eqnarray}
N(\varepsilon_F)= N_0(\varepsilon_F^{(0)}).
\label{nconserve}
\end{eqnarray}
%%%%%%%%%%%%%%%%%%%%%%%%%%%%%%%%%%
%
Here
%
%%%%%%%%%%%%%%%%%%%%%%%%%%%%%%%%%%
\begin{eqnarray}
N= \int_0^\infty \frac{\nu (\varepsilon)}{\exp{\left( \frac{\varepsilon-\varepsilon_F}{T} \right) }+1}d \varepsilon
\label{electrondensity}
\end{eqnarray}
%%%%%%%%%%%%%%%%%%%%%%%%%%%%%%%%%%
%
is the  density of  electrons moving under conditions of the
magnetic breakdown, having the density of states $\nu(\varepsilon)$  defined
by Eq.(\ref{density}) and $T$ is the temperature. We define
%
%%%%%%%%%%%%%%%%%%%%%%%%%%%%%%%%%%
\begin{eqnarray}
N_0(\varepsilon_F^{(0)}) =\nu_{in} \varepsilon^{(0)}_F
\label{nuInintial}
\end{eqnarray}
%%%%%%%%%%%%%%%%%%%%%%%%%%%%%%%%%%
%
as the density of electrons in the initial lattice (in the absence of the
DW) with the initial density of states $\nu_{in}= S_+^\prime /(\pi \hbar)^2$ and the initial Fermi energy $\varepsilon_F^{(0)}$ (for the dispersion law in Eq.(\ref{Bechgaard_dispersion}), $\nu_{in}= 2 b^\ast/v_F( \pi \hbar)^2$, see Eq.(\ref{S and S prime})).

In order to carry out the integration in Eq.(\ref{electrondensity}), we
use the Fourier expansion of DOS (see Appendix \ref{fourierapp}, Eq.(\ref{Fourier})).
Assuming that temperature  $T$ is sufficiently large with respect to $\hbar \omega_c$, we
keep only the first Fourier harmonics with $A_0=1$ and $A_2 =-|r|^2$ (see Eq.(\ref{A2})).
Inserting it in Eq.(\ref{electrondensity}) one finds
%
%%%%%%%%%%%%%%%%%%%%%%%%%%%%%%%%%%
\begin{eqnarray}
N= \nu_{in} \varepsilon_F -2 \nu_{in}\int_0^\infty \frac{|r(\varepsilon)|^2
\cos{S_+(\varepsilon})/\sigma}{\exp{(\varepsilon-\varepsilon_F)/T}+1} d\varepsilon.
\label{electrondensity2}
\end{eqnarray}
%%%%%%%%%%%%%%%%%%%%%%%%%%%%%%%%%%
%
While writing Eq.(\ref{electrondensity2}), we took into account the
dependence of the reflection probability at MB
regions $|r|^2$  on electron energy $\varepsilon$. As it is shown in
Appendix \ref{dynamics}, $r(\varepsilon)$ is equal to zero  at
$\varepsilon<\varepsilon_c^{(2)}$, where $\varepsilon_c^{(2)}
<\varepsilon_F^{(0)}$ is the energy at which the open electron
trajectories touch each other at the fixed DW wave vector (see Fig. \ref{configurations1}).
Taking the integral in Eq.(\ref{electrondensity2}), one finds
%
%%%%%%%%%%%%%%%%%%%%%%%%%%%%%%%%%%
\begin{eqnarray}
N= \nu_{in} \varepsilon_F - 4\pi \nu_{in}|r(\varepsilon_F)|^2 T \exp\left\{-\frac{\pi T}{\hbar \omega_c}\right\}\sin{\frac{S_+(\varepsilon_F)}{\sigma}}. \nonumber\\
\label{electrondensity3}
\end{eqnarray}
%%%%%%%%%%%%%%%%%%%%%%%%%%%%%%%%%%
%
Using Eq. (\ref{nconserve})  and Eq. (\ref{nuInintial}), one finds the
correction to the Fermi energy $\delta \varepsilon_F = \varepsilon_F
-\varepsilon_F^{(0)}$ in the form
%
%%%%%%%%%%%%%%%%%%%%%%%%%%%%%%%%%%
\begin{eqnarray}
\delta \varepsilon_F= |r|^2 4\pi T \exp{\left\{-\frac{\pi T}{
\hbar \omega_c} \right\}} \sin{ \frac{S_+}{\sigma}}.
\label{FermiCorrection}
\end{eqnarray}
%%%%%%%%%%%%%%%%%%%%%%%%%%%%%%%%%%
%
In the right-hand side of the above equation, all the quantities are
taken at $\varepsilon=\varepsilon_F^{(0)}$.
Using Eq.(\ref{FermiCorrection}) and neglecting terms of the order
of $|r|^2 \hbar \omega_c/\varepsilon_F^{(0)}$, one finds a correction
to the thermodynamical potential of electrons caused by the arising
of the DW in the form
%
%%%%%%%%%%%%%%%%%%%%%%%%%%%%%%%%%%
\begin{eqnarray}
(\delta \Omega)_{T,\varepsilon_F^{(0)}} =
N_0(\varepsilon_F^{(0)})|r|^2 4\pi T \exp\left\{-\frac{\pi T
}{\hbar \omega_c}\right\}\sin \frac{S_+}{\sigma}.
 \label{ThermPotential}
\end{eqnarray}
%%%%%%%%%%%%%%%%%%%%%%%%%%%%%%%%%%
%
Using Eq.(\ref{ThermPotential}), and taking into account that a
correction to the free energy $(\delta F)_{T,n}$ at constant $T$ and
$n$ is equal to $(\delta \Omega)_{T,\varepsilon_F^{(0)}}$ at
constant  $T$ and $\varepsilon_F^{(0)}$ (see, e.g., Ref.
\onlinecite{Landau1}), one finds change of the free energy per one particle
%
%%%%%%%%%%%%%%%%%%%%%%%%%%%%%%%%%%
\begin{eqnarray}
\delta F = |r|^2 4 \pi T\exp\left\{-\frac{\pi T
}{\hbar \omega_c}\right\} \sin \frac{S_+}{\sigma} + \hbar \omega_{\textbf{Q}}
\frac{\Delta^2}{2 g^2},
\label{FreeEnergy1}
\end{eqnarray}
%%%%%%%%%%%%%%%%%%%%%%%%%%%%%%%%%%
%
where the last term is the lattice elastic energy given by phonon
frequency $\omega$ at momentum $\textbf{Q}$, $g$ is the electron-phonon
coupling constant, and $\Delta$ is the energy gap produced in the
electron spectrum by the DW in the absence of the magnetic field
\cite{Gruener} (see Appendix \ref{dynamics} - section 1 for details of gap opening in the electron spectrum and DW potential matrix elements).

\begin{figure}
\centerline{\includegraphics[width=0.75\columnwidth]{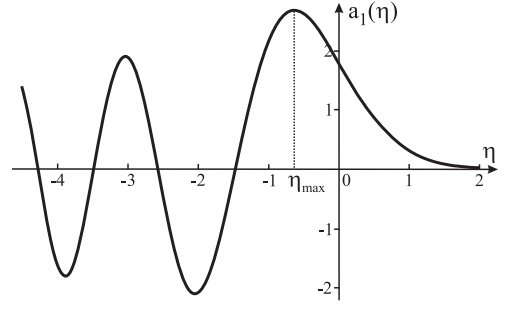}}
\caption{Coefficient $a_1(\eta)$ given by expression
(\ref{a1a2}) where $\eta_{max}$ denotes the position of global
maximum which determines optimal instability vector $Q_x$ via the
condition $\eta(Q_x, \varepsilon_F^{(0)}) = \eta_{max}$.}
\label{anu}
\end{figure}

\begin{figure}
\centerline{\includegraphics[width=0.85\columnwidth]{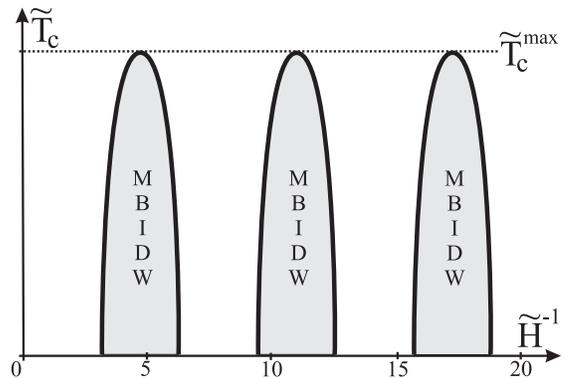}}
\caption{The phase diagram presenting the critical temperature of
MBIDW transition depending on inverse magnetic field based on
expression (\ref{Tc2}). The chosen scales are $\tilde{T}_c \equiv
\frac{\pi}{\hbar \omega_c}T_c$, $\tilde{H}^{-1}\equiv
\frac{cS_+}{\hbar e}H^{-1}$ and $\tilde{T}_c^{max}=\ln{\left(
\frac{56}{\beta^2} \lambda \left( \frac{\varepsilon_F}{\hbar \omega_c} \right)^{\frac{1}{3}}
\right)}$.}
\label{TH}
\end{figure}

As one can see from Eq.(\ref{FreeEnergy1}),  the electronic part of the
free energy is positive if $\sin S_+/\sigma >0$, meaning that the considered
MB induced Peierls transition may take place only at
magnetic fields for which $\sin S_+/\sigma < 0$.
Also, Eq. (\ref{FreeEnergy1}) shows that, due to the dependence of $|r|^2$ on the parameter $a_1$ (see Eq. (\ref{rsquared})) emerging from the quantum-mechanical solution of the MB problem through the band-touching region (see Appendix \ref{dynamics}-section 4 and Eq. (\ref{a1a2})), the largest (negative) energy gain for electrons is proportional to the largest value of
parameter $a_1$. On the other hand, parameter $a_1$ depends on the electron energy as well as on the band configuration defined by the DW momentum through parameter $\eta$ (the parameter $\eta$ is given by Eq. (\ref{parameters}), see also Fig. \ref{configurations1}),
%
%%%%%%%%%%%%%%%%%%%%%%%%%%%%%%%
\begin{eqnarray}{\label{a1}}
a_1 = 2^{2/3} \pi \textrm{Ai} \left( 2^{2/3} \eta \right)
\end{eqnarray}
%%%%%%%%%%%%%%%%%%%%%%%%%%%%%%%
%
where $\textrm{Ai}(x)$ is an Airy function.
As it follows from
Eq.(\ref{a1}), $a_1 \sim |\eta|^{-1/4}$ for $|\eta| \gg 1$, while
$a_1 \sim 1$ for $|\eta| \lesssim 1$ (see Fig. \ref{anu}).
Therefore, the deformation momentum that stabilizes the DW is found from
the condition $\eta (Q_x, \varepsilon_F^{(0)})= \eta_{max} \approx
-0.65$, and hence it is equal to
%
%%%%%%%%%%%%%%%%%%%%%%%%%%%%%%%%%%
\begin{equation}
Q_x = 2p_F \left\{1- \eta_{max} \left( \frac{\hbar
\omega_c}{\varepsilon_F}\right)^{\frac{2}{3}} \beta \right\}-\frac{2}{v_F} (2t_y -V_{+}),
\label{Q}
\end{equation}
%%%%%%%%%%%%%%%%%%%%%%%%%%%%%%%%%%
%
where $\beta \equiv 2 \left(\pi^2 t_y / t_x \right)^{1/3} \cos^{-1}{(ap_F/\hbar)}$ is a parameter appearing due to the anisotropy of the band (see Appendix \ref{dynamics} - section 1-2, Eq. (\ref{beta})), which is in the case of quater-filled Bechgaard salts of the order of 1.
The size of the pocket $S_{small}$ is also determined by $\eta_{max}$.
%
%%%%%%%%%%%%%%%%%%%%%%%%%%%%%%%%%%%%%%%%%%%%%%%%%%%%%%%%%%%%%%%%%%%%%%%%%%%%%
%%%%%%%%%%%%%%%%%%%%%%%%%%%%%%%%%%%%%%%%%%%%%%%%%%%%%%%%%%%%%%%%%%%%%%%%%%%%%
%
Taking this value of the deformation momentum $Q_x$ and using
Eq.(\ref{rsquared}), one finds that the probability of MB reflection is
%
%%%%%%%%%%%%%%%%%%%%%%%%%%%%%%%%%%
\begin{eqnarray}
|r|^2 = 7 \frac{\Delta^2}{\beta^2 (\hbar \omega_c)^{4/3}
\varepsilon_F^{2/3}}\Bigl( 1- \frac{\Delta^2}{\beta^2 (\hbar
\omega_c)^{4/3} \varepsilon_F^{2/3}}\Bigr),
\label{reflectance}
\end{eqnarray}
%%%%%%%%%%%%%%%%%%%%%%%%%%%%%%%%%%
%
where $\Delta$ is an order parameter related to the off-diagonal matrix element of the  DW potential, i.e. gap in the one-electron spectrum perturbed by the DW potential (see. Appendix \ref{dynamics}, section 1, Eq. (\ref{newenergies}), $\Delta \equiv |V_{12}|$).
Inserting it in Eq.(\ref{FreeEnergy1}) one obtains the free energy of
the system expanded in terms of the order parameter $\Delta$:
%
%%%%%%%%%%%%%%%%%%%%%%%%%%%%%%%%%%
\begin{eqnarray}
\delta F& =& \Bigl [\frac{28 \pi T\exp(-\pi T / \hbar \omega_c)\sin (S_+/\sigma)}{\beta^2 (\hbar \omega_c)^{4/3} \varepsilon_F^{2/3}}
+
\frac{\hbar \omega_{\textbf{Q}}}{2 g^2}\Bigr ] \Delta^2 \nonumber \\
&-&\frac{28 \pi T\exp(-\pi T / \hbar \omega_c) \sin
(S_+/\sigma)}{\beta^4 (\hbar \omega_c)^{8/3} \varepsilon_F^{4/3}} \Delta^4.
\label{FreeEnergy2}
\end{eqnarray}
%%%%%%%%%%%%%%%%%%%%%%%%%%%%%%%%%%
%
From here one finds the equation for the critical temperature  $T_c$
%
%%%%%%%%%%%%%%%%%%%%%%%%%%%%%%%%%%
\begin{eqnarray}
\frac{\exp(\pi T_c /\hbar \omega_c)}{(\pi T_c /\hbar \omega_c)}&=&
\frac{56 g^2}{\hbar \omega_{\textbf{Q}} \beta^2 (\hbar \omega_c)^{1/3}
\varepsilon_F^{2/3}} \left(-\sin{\frac{S_+}{\sigma}} \right) \nonumber \\
&\times &\Theta \left[-\sin{\frac{S_+}{\sigma}}\right].
\label{Tc1}
\end{eqnarray}
%%%%%%%%%%%%%%%%%%%%%%%%%%%%%%%%%%
%
Using Eq.(\ref{Tc1}) one may write  the free energy of the
system in terms of the order parameter $\Delta$ and the critical
temperature $T_c$ of MB induced Peierls transition  in the form
%
%%%%%%%%%%%%%%%%%%%%%%%%%%%%%%%%%%
\begin{equation}
\delta F =\frac{\hbar \omega_\textbf{Q}}{g^2} \Bigl\{ \frac{\pi
(T-T_c)}{\hbar \omega_c} \Delta ^2 + \frac{\Delta^4}{\beta^2 (\hbar
\omega_c)^{4/3}\varepsilon_F^{2/3}}\Bigr \}.
 \label{FreeEnergy2}
\end{equation}
%%%%%%%%%%%%%%%%%%%%%%%%%%%%%%%%%%
%
Based on the Eq.(\ref{Tc1}), the critical temperature can be approximately
written with the logarithmic accuracy as
%
%%%%%%%%%%%%%%%%%%%%%%%%%%%%%%%%%%
\begin{equation}
T_c \approx \frac{\hbar \omega_c}{\pi}\ln\left\{\frac{56}{\beta^2} \lambda
\Bigl(\frac{\varepsilon_F }{\hbar \omega_c}\Bigr)^{\frac{1}{3}}
\left(-\sin\frac{S_+}{\sigma} \right) \right\}
\label{Tc2}
\end{equation}
%%%%%%%%%%%%%%%%%%%%%%%%%%%%%%%%%%
%
provided the expression under the logarithm is greater than 1. Here
$\lambda =g^2/(\hbar \omega_{\textbf{Q}} \varepsilon_F)$ is the dimensionless
electron-phonon coupling parameter (see Ref. \onlinecite{Gruener}). The phase diagram of MBIDW, the
dependence of the critical temperature on the inverse magnetic field, is shown in Fig. \ref{TH}. Its peculiar behavior in the inverse magnetic field, exhibiting periodic appearance and disappearance of ordered phases within normal conducting phase, opens the possibility of magnetic filed controlled conducting state of the sample. It changes from the metallic state ($\tilde{T}_c =0$) to a nearly isolating one ($\tilde{T}_c \neq 0$), thus changing the sample resistance by orders of magnitude by a small variation of magnetic field.\\

Finally we compare the present result with our previous work \cite{ourwork}. Previously we have fixed the wave vector of the DW to $(2p_F,0)$, making electron trajectories simply to cross each other, in order to explore the possibility of MBIDW condensation in the simplest case even in a non-optimal configuration in which the MB regions have no internal structure. By letting the DW wave vector vary and finding the optimal value, we came to the new band configuration, in which the trajectories nearly touch each other, when the novel MB probability for tunneling region with particular internal structure had to be calculated. We note that the reflection probability $|r|^2 \approx \Delta^2 (\hbar \omega_c)^{-4/3} \varepsilon_F^{-2/3}$ in the present configuration is multiplied by additional large parameter $(\varepsilon_F / \hbar \omega_c)^{1/3}$ in comparison to the standard expression for magnetic breakdown $|r|_{std}^2 \approx \Delta^2 (\hbar \omega_c)^{-1} \varepsilon_F^{-1}$ used previously for the case of electron tunneling through MB regions without internal structure (see the details in Appendix \ref{dynamics} - Section 4, Eq. (\ref{rsquared})-(\ref{rsquaredstandard})). Therefore, one needs proportionally lower magnetic field to achieve the same effect. Indeed, the expression for the critical temperature in the present case (\ref{Tc2}) contains the same large parameter under logarithm which defines $T_c^{max}$ thus rising it proportionally.\\
Regarding the experimental observation of the effect, it depends a lot on the quality of the sample. The longitudinal electric fields in the sample, i.e dislocation fields, destroy quantum coherence of semiclassical wave packages thus deteriorating the MBIDW effect. The ad hoc criterion for the MB related effects, like the one that we predict, is usually the ability of the sample to exhibit de Haas van Alphen effect since it undergoes the same restrictions.
%
%
%%%%%%%%%%%%%%%%%%%%%%%%%%%%%%%%%%%%%%%%%%%%%%%%%%%%%%%%%%%%%%%%%%%%%%%%%%%%%
%%%%%%%%%%%%%%%%%%%%%%%%%%%%%%%%%%%%%%%%%%%%%%%%%%%%%%%%%%%%%%%%%%%%%%%%%%%%%

\section{conclusion}
We have shown that system with an open
Fermi surface under a homogeneous magnetic field $H$ is  unstable
with respect to a structural phase transition of a peculiar type, under
which an open Fermi surface is transformed into a chain of large
pockets separated with small ones having
a strong band dispersion and playing the role of
effective barriers between them. Quantum tunneling between
the neighboring large pockets caused by the magnetic field (magnetic breakdown) transforms
the electron spectrum into a series of alternating very narrow
energy gaps and bands which decreases the energy of the electronic
system that in its turn  stabilizes the density wave. We have
found that the optimal deformation momentum of the density wave
has a completely "anti-nesting" character at which the shifted
branches of the Fermi surface nearly touch each other. We have also
shown that the phase diagram containing the critical temperature of this magnetic breakdown
induced phase transition $T_c(H^{-1})$ is nearly periodic in the
inverse magnetic field, featuring a series of alternating narrow
"MBIDW-windows" ($T_c \neq 0$) and gaps ($T_c = 0$).
Comparing the present result with the previous one for simple $(2p_F,0)$ instability \cite{ourwork}, that generates MB tunneling between neighboring pockets without peculiar band structure inbetween, we find that now the MB tunneling probability in enhanced by large factor $(\varepsilon_F / \hbar \omega_c)^{1/3}$ consequently leading to the proportionally lower magnetic field and higher critical temperature for the same MBIDW phase transition.

\textbf{Acknowledgements.} A.K. gratefully acknowledges the hospitality of the University of Zagreb. The work is supported by project
119-1191458-1023 of Croatian Ministry of Science, Education and Sports.

\appendix

\section{Dynamics of electrons under magnetic field near points of the topological transition in the electron
spectrum.}\label{dynamics}

The conventional magnetic breakdown matrix \cite{Slutskin} describes the situation in which the electron tunnels between two large pockets of different electron energy bands (inter-band tunneling).
In this Appendix we analytically investigate the dynamics of electrons in the vicinity of the touching points of two  classical trajectories (see Fig. \ref{trajectories}) that is
near the topological phase transition of $2\frac{1}{2}$ order \cite{Lifshitz212}. In this case the magnetic breakdown  combines both the inter-band tunneling between neighboring large and small orbits and the intra-band one between neighboring large orbits that qualitatively changes the conventional magnetic breakdown matrix.

\subsection{The band structure in the absence of magnetic
field}\label{orbits}

In the two-dimensional case the electron energy band within the tight binding description can be written in the form
%
%%%%%%%%%%%%%%%%%%%%%%%%%%%%%%%%%%
\begin{eqnarray}
\varepsilon (\textbf{p})= - 2 t_x \cos\frac{a p_x}{\hbar} - 2 t_y \cos\frac{b p_y}{\hbar},
\label{initialBechgaard}
\end{eqnarray}
%%%%%%%%%%%%%%%%%%%%%%%%%%%%%%%%%%
%
where $\textbf{p}=(p_x,p_y)$ denotes electron momentum, $t_x$, $t_y>0$ are electron transfer integrals and $a$, $b$ are lattice constants in $x-$ and $y-$direction respectively with an energy origin taken in the middle of the band. Fermi surface, defined by $p_{x}^{F}(p_{y})$, follows from the condition $\varepsilon [p_{x}^{F}(p_{y}), p_{y}] =\varepsilon_{F}$, where the Fermi energy $\varepsilon_{F}$ is determined
by the number of available electrons filling the band. This surface includes two points $\pm p_{F} \equiv \pm p_{x}^{F}(p_{y}=\hbar\pi/2b)$, that are roots of the expression $\varepsilon_{F}= - 2 t_x \cos (a p_F / \hbar)$, where points $\pm p_{F}$ determine positions of two ("upper" and "lower") strict Fermi planes in the case $t_{y}=0$.\\

Now we assume $t_x \gg t_y$. This means that $p_{x}^{F}(p_y)$ will always be near
$p_{F}$ or $- p_{F}$, and that we can expand the first term in (\ref{initialBechgaard})
in terms of $p_{x}^{F}(p_y) \mp p_F$. With Fermi velocity given as
%
%%%%%%%%%%%%%%%%%%%%%%%%%%%%%%%%%%
\begin{equation}
v_F \equiv \frac{2 a t_x}{\hbar} \sin\frac{p_F a}{\hbar},
\label{Fermivelocity}
\end{equation}
%%%%%%%%%%%%%%%%%%%%%%%%%%%%%%%%%%
%
one gets two sheets ("$+$" - upper, always close to $+p_{F}$, "$-$" - lower, always close to $-p_{F}$,
in accordance with Fig. \ref{trajectories}) representing the open Fermi surface:
%
%%%%%%%%%%%%%%%%%%%%%%%%%%%%%%%%%%
\begin{equation}
p_{x}^{F,\pm}(p_y) =\pm p_F \pm \frac{2 t_y}{v_F} \cos \frac{b p_y}{\hbar}.
\label{sheets}
\end{equation}
%%%%%%%%%%%%%%%%%%%%%%%%%%%%%%%%%%
%
The "iso-energetic" sheets for energies $\varepsilon$ around $\varepsilon_{F}$ are given
by
%
%%%%%%%%%%%%%%%%%%%%%%%%%%%%%%%%%%
\begin{equation}
p_{x}^{\pm}(p_y) = \pm p_{F} \pm  \frac{(\varepsilon -\varepsilon_F)}{v_F} \pm  \frac{2 t_y}{v_F} \cos \frac{b p_y}{\hbar}.
\label{iso-energetic-sheets}
\end{equation}
%%%%%%%%%%%%%%%%%%%%%%%%%%%%%%%%%%
%
Without periodic perturbation, the momentum $\textbf{p}$, i.e. each pair $(p_x, p_y)$ of its components, is
a good quantum number. We denote the wave functions by $\varphi_{\textbf{p}}^{\pm}(\textbf{r})$. The corresponding eigen-energies are
%
%%%%%%%%%%%%%%%%%%%%%%%%%%%%%%%%%%
\begin{equation}
\varepsilon_{\pm}(\textbf{p}) = \varepsilon_F \pm v_F (p_x \mp p_F) - 2t_y cos\frac{b p_y}{\hbar}
\label{energies}
\end{equation}
%%%%%%%%%%%%%%%%%%%%%%%%%%%%%%%%%%
%
for $p_x$ close to $\pm p_F$.

\begin{figure}
\centerline{\includegraphics[width=8.0cm]{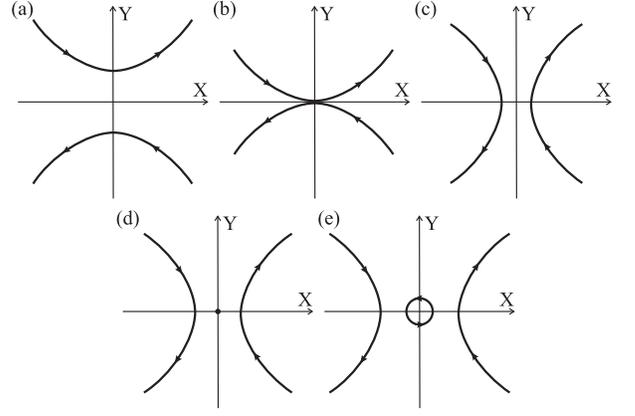}}
\caption{Possible electron band configurations following from Eq.
(\ref{newenergies}) given in the form $X^2=\delta\varepsilon \mp
\sqrt{Y^2+|V_{12}|^2}$ where $\delta\varepsilon \equiv \varepsilon_0 + V_+ - \varepsilon$, $X\equiv \sqrt{\alpha}p_y$, $Y\equiv vp_x+V_-$
(definitions of used constants are given in Appendix \ref{dynamics}). Depending
on electron energy $\varepsilon$, there are several characteristic
band configurations: (a) $\varepsilon<\varepsilon_c^{(2)}$, (b)
$\varepsilon=\varepsilon_c^{(2)}$, (c)
$\varepsilon_c^{(2)}<\varepsilon<\varepsilon_c^{(1)}$, (d)
$\varepsilon=\varepsilon_c^{(1)}$, (e)
$\varepsilon>\varepsilon_c^{(1)}$, where
$\varepsilon_c^{(1,2)}=\varepsilon_0+V_+\pm |V_{12}|$. Arrows denote
electron trajectories in the presence of magnetic field.}
\label{configurations1}
\end{figure}

In the presence of a density wave, the Schr\"{o}dinger
equation with perturbation is
%
%%%%%%%%%%%%%%%%%%%%%%%%%%%%%%%%%%
\begin{equation}\label{H_tot}
\left( \hat{H}_0 + V(x) \right) \Psi(\textbf{r}) = E \Psi(\textbf{r}),
\end{equation}
%%%%%%%%%%%%%%%%%%%%%%%%%%%%%%%%%%
%
where $\hat{H}_0$ is the  Hamiltonian of a conduction electron in the unperturbed crystal discussed above.
As assumed in the main text, the DW modulation potential $V(x)$, periodic with momentum $\textbf{Q}$, shifts the branches of the Fermi surface and forms the touching points (see Fig. \ref{trajectories}) which are the subject of analysis in this Appendix. Since the perturbation couples states from "$+$" and "$-$" sheets of the Fermi surface, $Q_x$ should be close to $2p_F$, deviating from it on the scale $t_y/v_F$. $V(x)$ is
presumably dependent on $x$ only, also it is relatively weak with respect to the bandwidth parameters $v_F p_F$ and $t_y$, therefore it can be treated as a perturbation to the Hamiltonian $\hat{H}_0$ outside the range of these touching points. Within this range we perform
the standard diagonalization procedure for two (almost) degenerate states. The corresponding wave functions
for these states are $\varphi_{1} \equiv \varphi^{-}_{p_x^{(1)},p_y}(\textbf{r}) $ and
$\varphi_{2} \equiv \varphi^{+}_{p_x^{(2)},p_y}(\textbf{r})$, where $p_x^{(1,2)}$ are defined in the main text, Eq. (\ref{pshift}).
The sought-for wave functions $\Psi$ near the degeneracy points in Fig. \ref{trajectories} can be written as linear combinations
%
%%%%%%%%%%%%%%%%%%%%%%%%%%%%%%%%%%
\begin{eqnarray}
\Psi = \beta_1 \varphi_1 + \beta_2 \varphi_2.
\label{perturbfunction}
\end{eqnarray}
%%%%%%%%%%%%%%%%%%%%%%%%%%%%%%%%%%
%
Inserting $\Psi$ into Eq.(\ref{H_tot}), multiplying the resulting equation by $ \varphi_1^\ast$ and then by
$ \varphi_2^\ast$ and integrating, one obtains the following set of equations for two branches $s=1,2$ in the the energy spectrum,
$E_s(\textbf{p})$, and corresponding coefficients $\beta_s$:
%
%%%%%%%%%%%%%%%%%%%%%%%%%%%%%%%%%%
\begin{eqnarray}{\label{system0}}
\left( \varepsilon_1 (\textbf{p}) + V_{11} - E \right) \beta_1(\textbf{p}) + V_{12} \beta_2(\textbf{p}) = 0 \nonumber \\
V_{12}^* \beta_1(\textbf{p}) + \left( \varepsilon_2 (\textbf{p}) +
V_{22} - E \right) \beta_2(\textbf{p}) = 0.
\end{eqnarray}
%%%%%%%%%%%%%%%%%%%%%%%%%%%%%%%%%%
%
Here we define $\varepsilon_1 \equiv \varepsilon_{-}(p_x-Q_x, p_y)$,
$\varepsilon_2 \equiv \varepsilon_{+}(p_x, p_y)$ and $V_{ss'} \equiv \langle
\varphi_s | V(x) | \varphi_{s'} \rangle$ are matrix
elements of the perturbing potential. Using the spectrum (\ref{energies}), after a convenient shift of momentum origin $p_x \rightarrow p_x+Q_x/2$, $p_y \rightarrow p_y+b^\ast/2$ (to the trajectory-touching point), we obtain
%
%%%%%%%%%%%%%%%%%%%%%%%%%%%%%%%%%%
\begin{equation}{\label{bands0}}
\varepsilon_{1,2} (\textbf{p}) = \varepsilon_0 \mp v_F p_x - 2t_y \left( 1- \cos{\frac{b p_y}{\hbar}} \right),
\end{equation}
%%%%%%%%%%%%%%%%%%%%%%%%%%%%%%%%%%
%
where
%
%%%%%%%%%%%%%%%%%%%%%%%%%%%%%%%%%%
\begin{equation}
\varepsilon_0 \equiv \varepsilon_F - v_F p_F +v_F Q_x/2+2t_y.
\label{epsilon0}
\end{equation}
%%%%%%%%%%%%%%%%%%%%%%%%%%%%%%%%%%
%
From Eq.(\ref{system0}) it follows that the eigenvalues of the total
Hamiltonian $\hat{H}$ (Eq. (\ref{H_tot})) in the vicinity of the  degeneracy points
are
%
%%%%%%%%%%%%%%%%%%%%%%%%%%%%%%%%%%
\begin{equation}\label{newenergies}
E_{1,2}(\textbf{p})=\varepsilon_0 + V_+ - \alpha p_y^2 \mp
\sqrt{(v_F p_x+V_-)^2+|V_{12}|^2},
\end{equation}
%%%%%%%%%%%%%%%%%%%%%%%%%%%%%%%%%%
%
with
%
%%%%%%%%%%%%%%%%%%%%%%%%%%%%%%%%%%
\begin{equation}
\alpha\equiv \frac{t_y b^2}{\hbar^2}=\frac{1}{2m_y^\ast}, \hspace{0.5cm} V_{\pm} \equiv \frac{1}{2}(V_{22}\pm V_{11}),
\label{alpha}
\end{equation}
%%%%%%%%%%%%%%%%%%%%%%%%%%%%%%%%%%
%
$m_y^\ast$ has the role of a ”dynamic” effective electron mass in the transverse direction, obtained after expanding
$\varepsilon_{1,2} (\textbf{p})$ near the touching point $(p_x,p_y)= (p_F - 2t_y/v_F,b^\ast / 2)$ (now shifted to the momentum space origin), i.e.
$\varepsilon_{1,2}\approx \varepsilon_0 \mp v_F p_x-\alpha p_y^2$.
The eigenvalues
$E_{1,2}(\textbf{p})$ are two new electron energy bands separated by
an energy gap $\Delta = |V_{12}|$. An investigation of
Eq.(\ref{newenergies}) shows that there are two peculiar points
$\varepsilon_c^{(1,2)}=\varepsilon_0+V_+\pm |V_{12}|$ in the new
electron dispersion law  at which the equipotential surfaces
$E_{1,2}(p_x,p_y) =\varepsilon$ change their topology (see
Fig.\ref{configurations1}).
%
%%%%%%%%%%%%%%%%%%%%%%%%%%
%%%%%%%%%%%%%%%%%%%%%%%%%%%%

\subsection{Magnetic breakdown through the band-touching region}

The Hamiltonian of a conduction electron in a metal under magnetic
field  in the coordinate representation is $\hat{H}(\hat{\textbf{p}} -
\frac{e}{c}\textbf{A},\textbf{ r})$, where
$\hat{H}(\hat{\textbf{p}},\textbf{ r})$ is the electron Hamiltonian
in the absence of magnetic field,  $\textbf{p} =-i\hbar
\partial / \partial \textbf{r}$ is the electron \emph{momentum} operator
and $\textbf{A}$ is the vector potential.

In the momentum presentation, I.M. Lifshitz and L.
Onsager \cite{Lifshits} suggested the conduction electron Hamiltonian in the form
Eq.(\ref{LO}), in which the electron \emph{quasi-momentum}
$\textbf{p}$ in the electron dispersion law
$\varepsilon_s(\textbf{p})$ is substituted with $\textbf{p}-
(e/c)\hat{\textbf{A}}$, where  the gauge potential is chosen in the Landau gauge, i.e. $\hat{\textbf{A}}
= (-H \hat{y}, 0,0), \; \hat{y} = i \hbar \partial /\partial p_y$.
As it was proved by Zilberman \cite{Zilberman}, one gets the Onsager-Lifshitz
Hamiltonian, in the case when the band number is conserved, by using the
complete orthonormal set of modified Bloch functions
$\chi_{s,\textbf{p}}(\textbf{r})= e^{i\textbf{p}\textbf{r}}
u_{s,p_x+eHy/c,p_y}(\textbf{r})$, where $u_{s,p_x,p_y}(\textbf{r})$ is the
periodic function. Such $\chi_{s,\textbf{p}}(\textbf{r})$ are called the Zilberman functions.

As it was shown in Refs. \onlinecite{SK,S}, in the region of
magnetic breakdown, where the band number is not conserved,  one gets
the Hamiltonian by expanding the electron wave function $\Psi
(\textbf{r})$ in the form $\Psi(\textbf{r})=\sum_{s=1}^{2} \int
d\textbf{p} g_s(\textbf{p}) \chi_{s,\textbf{p}}(\textbf{r})$
resulting in the above-mentioned substitution in Eq.(\ref{system0}):
%
%%%%%%%%%%%%%%%%%%%%%%%%%%%%%%%%%%
\begin{eqnarray}{\label{systemH1}}
\left( \varepsilon_1 (P_{x0}+i\sigma \frac{d}{dp_y},p_y) + V_{11} - E \right) g_1(\textbf{p}) + V_{12} g_2(\textbf{p}) = 0 \nonumber \\
V_{12}^* g_1(\textbf{p}) + \left( \varepsilon_2 (P_{x0}+i\sigma
\frac{d}{dp_y},p_y) + V_{22} - E \right) g_2(\textbf{p}) = 0 \nonumber\\
\end{eqnarray}
%%%%%%%%%%%%%%%%%%%%%%%%%%%%%%%%%%
%
where $\sigma\equiv e\hbar H /c$, $V_{ss'}=\langle \chi_s | V(\textbf{r})
| \chi_{s'} \rangle$ and $P_{x0}$ is conserved longitudinal electron
momentum due to the chosen Landau gauge. Introducing expression (\ref{bands0}) and expansion
around the touching point as in previous section, then summing and subtracting the equations in
(\ref{systemH1}), lead us to system
%
%%%%%%%%%%%%%%%%%%%%%%%%%%%%%%%%%%
\begin{eqnarray}{\label{systemH2}}
iv_F \sigma \frac{dg^{(-)}}{dp_y} +(-\alpha p_y^2 + \varepsilon_c^{(1)}-E)g^{(+)} = 0 \nonumber \\
iv_F \sigma \frac{dg^{(+)}}{dp_y} +(-\alpha p_y^2 +\varepsilon_c^{(2)}-E)g^{(-)} = 0,
\end{eqnarray}
%%%%%%%%%%%%%%%%%%%%%%%%%%%%%%%%%%
%
with $V_{12}=|V_{12}|e^{i\theta}$ and $g^{(\pm)}=\bar{g}_2 \pm
\bar{g}_1$, $\bar{g}_{1,2}(\textbf{p})\equiv
\exp[-\frac{i}{\sigma}(P_{x0}+\frac{V_-}{v})p_y \mp
i\frac{\theta}{2}] g_{1,2}(\textbf{p})$. Further we substitute
$p_y\equiv m_y^*v_F \kappa^{1/3}\xi$ and introduce auxiliary functions
$\textbf{u}(\xi)=(u_1,u_2)$ related to $g^{(\pm)}$ as
$g^{(\pm)}(\xi)=\pm u_1(\xi)\exp{i(\xi^3/3+\eta\xi)} +
u_2(\xi)\exp{-i(\xi^3/3+\eta\xi)}$. System (\ref{systemH2}) reduces to
%
%%%%%%%%%%%%%%%%%%%%%%%%%%%%%%%%%%
\begin{eqnarray}{\label{systemH3}}
i\frac{du_1(\xi)}{d\xi}&=&-\gamma e^{-i\left(\frac{2}{3}\xi^3 + 2\eta\xi\right)} u_2(\xi) \nonumber \\
i\frac{du_2(\xi)}{d\xi}&=&\gamma e^{i\left(\frac{2}{3}\xi^3 +
2\eta\xi\right)} u_1(\xi),
\end{eqnarray}
%%%%%%%%%%%%%%%%%%%%%%%%%%%%%%%%%%
%
where $\omega_y \equiv
\frac{eH}{m_y^\ast c}$, $\kappa \equiv \hbar \omega_y / \varepsilon_b$, while
$\varepsilon_b=m_y^\ast v_F^2/2$ is of the order of Fermi energy, and two
emerging parameters are defined as
%
%%%%%%%%%%%%%%%%%%%%%%%%%%%%%%%%%%
\begin{eqnarray}{\label{parameters}}
\gamma &\equiv& -\frac{|V_{12}|}{\beta (\hbar \omega_c)^{2/3}\varepsilon_F^{1/3}} \nonumber \\
\eta &\equiv& -\frac{\varepsilon_0 + V_+ -\varepsilon}{\beta (\hbar
\omega_c)^{2/3}\varepsilon_F^{1/3}},
\end{eqnarray}
%%%%%%%%%%%%%%%%%%%%%%%%%%%%%%%%%%
%
with
%
%%%%%%%%%%%%%%%%%%%%%%%%%%%%%%%%%%
\begin{equation}{\label{beta}}
\beta \equiv \left(\frac{\omega_y}{\omega_c}\right)^{\frac{2}{3}} \left(\frac{\varepsilon_b}{\varepsilon_F}\right)^{\frac{1}{3}}=
\left(\frac{\pi^2 t_y}{t_x}\right)^{\frac{1}{3}}\frac{2}{\cos{(ap_F/\hbar})}.
\end{equation}
%%%%%%%%%%%%%%%%%%%%%%%%%%%%%%%%%%
%
In the case of quater-filled Bechgaard salts parameter $\beta \approx 0.7$ is of the order of 1.
Parameter $\gamma$, that mixes functions $u_1$ and $u_2$,  appears as
a magnetic breakdown tunneling parameter, while $\eta$ gives
criterion of validity of semiclassical description, i.e. as long as
$|\eta| \lesssim 1$ and $|\xi| \lesssim 1$, the full quantum treatment is
required because the functions in Eq.(\ref{systemH3}) are not fast
oscillating. We should note the difference between this case and
conventional magnetic breakdown problem in metals when semiclassical
approximation is valid as long as the area of closed electron orbit
in $p$-space is much bigger than $\sigma$. Here much wider area than
just small closed electron orbit in Fig.\ref{configurations1}(e)
requires the quantum treatment. Parameter $|\gamma| \ll 1$
is required to be small in our approach consequently permitting us
to build a perturbation theory $\textbf{u}=\textbf{u}^{(0)}+\gamma
\textbf{u}^{(1)} + \gamma^2 \textbf{u}^{(2)} + ...$ in order to solve the
system (\ref{systemH3}). The asymptotic boundary conditions, i.e.
the requirement of matching $\textbf{u}(\xi)$ to semiclassical
solutions in the limit $\xi \rightarrow \pm \infty$, will be
determined in the next section. Integrating the system
(\ref{systemH3}) we get $\textbf{u}({-\infty}) = \textbf{u}^{(0)}$ and
$\textbf{u}({\infty}) = \hat{\mathrm{T}} \textbf{u}^{(0)}$, where
$\hat{\mathrm{T}}$ is unitary matrix with elements
$\mathrm{T}_{11}=1+\gamma^2a_2$, $\mathrm{T}_{12}=i\gamma a_1$,
with coefficients given by
%
%%%%%%%%%%%%%%%%%%%%%%%%%%%%%%%%%%
\begin{eqnarray}{\label{a1a2}}
a_1 &\equiv& \int_{-\infty}^{\infty} e^{-i(2x^3/3+2\eta x)}dx = 2^{2/3} \pi \textrm{Ai} \left( 2^{2/3} \eta \right) \\
a_2 &\equiv& \int_{-\infty}^{\infty} e^{-i(2x^3/3+2\eta x)}dx \int_{-\infty}^{x} e^{-i(2y^3/3+2\eta y)}dy, \nonumber
\end{eqnarray}
%%%%%%%%%%%%%%%%%%%%%%%%%%%%%%%%%%
%
where $\textrm{Ai}(x)$ is an Airy function. In terms of $\textbf{u}(\infty)$ we obtain
%
%%%%%%%%%%%%%%%%%%%%%%%%%%%%%%%%%%
\begin{equation}{\label{gbars}}
\bar{g}_{1,2}(\xi) \approx u_{1,2}(\infty) e^{\pm i\left(
\frac{\xi^3}{3} + \eta \xi \right)}
\end{equation}
%%%%%%%%%%%%%%%%%%%%%%%%%%%%%%%%%%
%
in the limit $|\xi| \gg 1$. From expressions (\ref{gbars}), one
easily gets starting coefficients
%
%%%%%%%%%%%%%%%%%%%%%%%%%%%%%%%%%%
\begin{equation}{\label{gstart}}
g_{1,2} \approx \frac{1}{\sqrt{v_F}} u_{1,2}(\infty) e^{i\left(
\frac{1}{\sigma}P_{x0}p_y \pm \frac{\theta}{2}\right)} e^{\pm
i\left( \frac{\xi^3}{3} + \eta_{1,2} \xi \right)}
\end{equation}
%%%%%%%%%%%%%%%%%%%%%%%%%%%%%%%%%%
%
with redefined parameters
%
%%%%%%%%%%%%%%%%%%%%%%%%%%%%%%%%%%
\begin{equation}{\label{etanew}}
\eta_s \equiv -\frac{\varepsilon_0 + V_{ss} - \varepsilon}{\beta (\hbar
\omega_c)^{2/3}\varepsilon_F^{1/3}}
\end{equation}
%%%%%%%%%%%%%%%%%%%%%%%%%%%%%%%%%%
%
into which we have absorbed dependency on $V$-matrix elements.
Factor $v_F^{-1/2}$ additionally appears due to normalization of
electron current density to 1. Note that $|\xi| \gg 1$ corresponds
to $m_y^\ast v_F \kappa^{1/3} \ll |p_y| \ll m_y^\ast v_F$ where it is possible to
expand the transversal electron dispersion up to square contribution
$\alpha p_y^2$ used in the equations above. That limit gives the
quantum mechanical description of the region in which the quantum MB
transitions are absent, but which still overlaps with the semiclassical
description (next section). There the matching of quantum and
semiclassical solution is done.

\subsection{The semiclassical solution}

Since magnetic breakdown transitions are absent in the region $|\xi|
\gg 1$, where $\textbf{u}(\xi)\approx const.$, one may use
Onsager-Lifshitz Hamiltonian in the form
%
%%%%%%%%%%%%%%%%%%%%%%%%%%%%%%%%%%
\begin{equation}\label{LOHamiltonian}
E_s\left( P_{x0} + i\sigma \frac{d}{dp_y}, p_y \right)
G_s(\textbf{p}) = \varepsilon G_s(\textbf{p}),
\end{equation}
%%%%%%%%%%%%%%%%%%%%%%%%%%%%%%%%%%
%
where electron dispersions attain the form $E_{s}(\textbf{p})
\approx \varepsilon_s(\textbf{p}) + V_{ss}$ since we are far away
from the band-touching point. Using the form of
$\varepsilon_{1,2}(\textbf{p})$ given by expression (\ref{bands0}),
one gets system of equations for $G_{1,2}$
%
%%%%%%%%%%%%%%%%%%%%%%%%%%%%%%%%%%
\begin{eqnarray}\label{LOsystem}
i\sigma v_F \frac{dG_{1,2}}{dp_y}=\left[ -v_F P_{x0} \pm \left( \varepsilon_0 + V_{11,22} \right. \right. \nonumber\\
\left. \left.- \varepsilon - 2t_y \left( 1-\cos{\frac{b p_y}{\hbar}} \right)
\right) \right] G_{1,2}.
\end{eqnarray}
%%%%%%%%%%%%%%%%%%%%%%%%%%%%%%%%%%
%
Equations in system (\ref{LOsystem}) are not coupled and integration
simply gives
%
%%%%%%%%%%%%%%%%%%%%%%%%%%%%%%%%%%
\begin{equation}\label{semiclsolution0}
G_{1,2}(\textbf{p})=\frac{C_{1,2}}{\sqrt{v_F}} e^{i\frac{1}{\sigma}
\left( P_{x0}p_y \mp S_{1,2}(p_y) \right)},
\end{equation}
%%%%%%%%%%%%%%%%%%%%%%%%%%%%%%%%%%
%
where $S_{s}(p_y)=(\varepsilon_0-2t_y+V_{ss}-\varepsilon)\frac{p_y}{v_F}+\frac{2\hbar
t_y}{b v_F}\sin{\frac{b p_y}{\hbar}}$ is semiclassical action,
$C_{1,2}$ are normalization constants and $v_F^{-1/2}$ factor again
appears due to normalization of electron current density to 1. In
order to match the obtained semiclassical solutions to the
quantum-mechanical solutions from previous chapter, we introduce
variable $\xi$ in the same way $p_y\equiv m_y^\ast v_F \kappa^{1/3}\xi$.
Also, since $m_y^\ast v_F b/\hbar \ll 1$, we expand $\sin{\frac{b
p_y}{\hbar}}$ in $S_s(p_y)$ within the semiclassical region $1 \ll
|\xi| \ll \kappa^{-1/3}$ into Taylor series up to $\sim \xi^3$. This
yields semiclassical solutions in the form
%
%%%%%%%%%%%%%%%%%%%%%%%%%%%%%%%%%%
\begin{equation}\label{semiclsolution}
G_{1,2}(\textbf{p})=\frac{C_{1,2}}{\sqrt{v_F}}
e^{i\frac{1}{\sigma}P_{x0}p_y} e^{\pm i\left( \frac{\xi^3}{3} +
\eta_{1,2} \xi \right)}.
\end{equation}
%%%%%%%%%%%%%%%%%%%%%%%%%%%%%%%%%%

\subsection{Magnetic breakdown tunneling matrix}\label{matrix}

To match quantum to semiclassical solution, i.e. to match the
coefficients $g_{1,2}$ to $G_{1,2}$, one more step is required. We
should note that Eq. (\ref{LOHamiltonian}) for coefficients
$G_{1,2}$ is obtained using expansion of solution
$\Psi(\textbf{r})=\sum_{m=1}^{2} \int d\textbf{p} G_m(\textbf{p})
X_{m,\textbf{p}}(\textbf{r})$ in terms of "total" Zilberman
functions $X_{m,\textbf{p}}(\textbf{r})= e^{i\textbf{p}\textbf{r}}
U_{m,p_x+eHy/c,p_y}(\textbf{r})$, $U(\textbf{r})$ is periodic, i.e.
featuring the total Hamiltonian (\ref{H_tot}) with perturbation
$V(\textbf{r})$. On the other hand, quantum solution, resulting with
coefficients $g_{1,2}$, is obtained using the expansion in terms of
Zilberman functions $\chi_{s,\textbf{p}}(\textbf{r})$ corresponding
to unperturbed Hamiltonian $\hat{H}_0$. In order to match two sets
of coefficients, the expansion in the same set of Zilberman
functions should be used, i.e. we expand $X$ in terms of $\chi$ as
$X_{m,\textbf{p}}(\textbf{r}) = \sum_{s=1}^{2} \beta_{s,m} \left(
P_{x0} + \frac{eHy}{c},p_y \right) \chi_{s,\textbf{p}}(\textbf{r})$.
Comparison of two expansions for $\Psi$ leads to connection between
coefficients $g_s(\textbf{p})=\sum_{m=1}^{2} \beta_{s,m} \left(
P_{x0} + \frac{eHy}{c},p_y \right) G_m(\textbf{p})$. After utilizing
the standard substitution as before $\frac{eHy}{c} \rightarrow
i\sigma \frac{d}{dp_y}$ for deriving the Onsager-Lifshitz
Hamiltonian, the previous expression reduces to set of differential
equations for coefficients $G_m$ in terms of given $g_s$, i.e.
%
%%%%%%%%%%%%%%%%%%%%%%%%%%%%%%%%%%
\begin{equation}\label{difeqG}
g_s(\textbf{p}) = \sum_{m=1}^2 \beta_{s,m}\left( P_{x0} + i\sigma
\frac{d}{dp_y},p_y \right) G_m(\textbf{p}).
\end{equation}
%%%%%%%%%%%%%%%%%%%%%%%%%%%%%%%%%%
%
However, being in the semiclassical limit, we use the form of
$G_{1,2}$ given by expression (\ref{semiclsolution0}) which, after
neglecting the terms $V_{ss}/v_F \ll |dS/dp_y| \sim p_F$, $S \equiv (\varepsilon_0-2t_y-\varepsilon)p_y/v_F$, reduces
system (\ref{difeqG}) to the set of algebraic equations
%
%%%%%%%%%%%%%%%%%%%%%%%%%%%%%%%%%%
\begin{equation}\label{algeqG}
g_s(\textbf{p}) = \beta_{s,1}\left(\frac{dS}{dp_y},p_y \right)
G_1(p_y) + \beta_{s,2}\left(-\frac{dS}{dp_y},p_y \right) G_2(p_y).
\end{equation}
%%%%%%%%%%%%%%%%%%%%%%%%%%%%%%%%%%
%
Coefficients $\beta_{sm}$ are easily obtained from the system
(\ref{system0}) and after normalization they read:
$\beta_{11}\approx \sqrt{|V_{12}|/V_{12}^*}$, $\beta_{22}\approx -i
\sqrt{V_{12}^*/|V_{12}|}$, $\beta_{12}=\beta_{21}\approx
|V_{12}|/(\varepsilon_2-\varepsilon_1) \ll 1$. After
neglecting $\beta_{12} \ll 1$ contributions in expression
(\ref{algeqG}), we obtain expressions for coefficients
%
%%%%%%%%%%%%%%%%%%%%%%%%%%%%%%%%%%
\begin{equation}\label{gG}
g_1(\textbf{p}) \approx e^{i\frac{\theta}{2}}G_1(\textbf{p}),
\hspace{0.5cm} g_2(\textbf{p}) \approx
-ie^{-i\frac{\theta}{2}}G_2(\textbf{p})
\end{equation}
%%%%%%%%%%%%%%%%%%%%%%%%%%%%%%%%%%
%
$(e^{i\theta}=V_{12}/|V_{12}|)$ valid both in "left"
($\xi\rightarrow -\infty$) and "right" ($\xi\rightarrow \infty$)
semiclassical region. Using the expression (\ref{gG}) we can match
coefficients given by Eqs. (\ref{gstart}) and
(\ref{semiclsolution}). Denoting the semiclassical normalization
constants $C_s$ from expression (\ref{semiclsolution}) in "left"
region as $C_s^{(I)}$ and in "right" region as $C_s^{(II)}$, the
matching conditions read
%
%%%%%%%%%%%%%%%%%%%%%%%%%%%%%%%%%%
\begin{eqnarray}\label{matchingsys}
C_1^{(I)} &=& u_1^{(0)} \\
-iC_2^{(I)} &=& u_2^{(0)} \nonumber\\
C_1^{(II)} &=& (1+\gamma^2 a_2)u_1^{(0)}+i\gamma a_1 u_2^{(0)} \nonumber\\
-iC_2^{(II)} &=& (1+\gamma^2 a_2^\ast)u_2^{(0)}-i\gamma a_1^\ast u_1^{(0)}.
\nonumber
\end{eqnarray}
%%%%%%%%%%%%%%%%%%%%%%%%%%%%%%%%%%
%
Finally, from the system (\ref{matchingsys}), we obtain connection
between the incoming and outgoing waves through the MB region (see
Fig. \ref{MBmatrix1})
%
%%%%%%%%%%%%%%%%%%%%%%%%%%%%%%%%%%
\begin{equation}\label{inout}
\left( \begin{array}{c}
C_1^{(I)} \\
C_2^{(II)} \end{array} \right) = \hat{\mathbf{\tau}} \left(
\begin{array}{c}
C_1^{(II)} \\
C_2^{(I)} \end{array} \right)
\end{equation}
%%%%%%%%%%%%%%%%%%%%%%%%%%%%%%%%%%
%
given in terms of MB tunneling matrix
%
%%%%%%%%%%%%%%%%%%%%%%%%%%%%%%%%%%
\begin{equation}\label{MBmatrix}
\hat{\mathbf{\tau}} = e^{i\Theta} \left( \begin{array}{cc}
t & r \\
-r^* & t \end{array} \right)
\end{equation}
%%%%%%%%%%%%%%%%%%%%%%%%%%%%%%%%%%
%
where
%
%%%%%%%%%%%%%%%%%%%%%%%%%%%%%%%%%%
\begin{equation}\label{tunnprobabilities}
t=1-\frac{1}{2}\gamma^2 |a_1|^2, \hspace{0.5cm} r=-\gamma a_1
\end{equation}
%%%%%%%%%%%%%%%%%%%%%%%%%%%%%%%%%%
%
define the tunneling probabilities of transmission through and
reflection at the MB region respectively. The corresponding phase is
$\Theta=-\gamma^2 \mathrm{Im}(a_2)$, all given up to $~\gamma^2$
accuracy. The main result of the paper is expressed in terms of
$|r|^2$ which we need up to $~\gamma^4$ accuracy. After a tedious,
but quite straightforward procedure analogous to the one presented
above, one obtains
%
%%%%%%%%%%%%%%%%%%%%%%%%%%%%%%%%%%
\begin{equation}\label{rsquared}
|r|^2 \approx |a_1|^2 \gamma^2 (1-\gamma^2),
\end{equation}
%%%%%%%%%%%%%%%%%%%%%%%%%%%%%%%%%%
%
where $a_1$ and $\gamma$ are defined by Eq.(\ref{a1a2}) and
Eq.(\ref{parameters}) respectively.
We  have obtained Eq.(\ref{rsquared}) by solving  the set of equations
Eq.(\ref{systemH1}) in the region where the semiclassical
approximation is not valid, using the perturbation theory in $|\gamma|
\ll 1$ and then using an asymptotic form of this solution to  match
semiclassical wave functions on the neighboring large orbits. Following from Eq.(\ref{a1a2}), one gets $a_1 \sim 1$ for $|\eta| \sim 1 $ and
%
%%%%%%%%%%%%%%%%%%%%%%%%%%%%%%%%%%
\begin{equation}\label{rsquared1}
|r|^2 \approx \frac{|V_{12}|^2}{\beta^2 (\hbar
\omega_c)^{4/3}\varepsilon_F^{2/3}}.
\end{equation}
%%%%%%%%%%%%%%%%%%%%%%%%%%%%%%%%%%

Also, one can see from Eqs.(\ref{systemH3}) and (\ref{parameters}), if
$|\eta| \gg 1 $ (achievable either by increasing the energy or further increase of DW vector $Q_x$), the small orbits in Fig. \ref{configurations1}(e) become semiclassically large and magnetic breakdown occurs in the small areas
$\sim \sigma$ \emph{between} them and the large neighboring orbits. However,
for $|\eta| \gg 1$, one has $a_1 \propto  |\eta|^{-1/4}$, consequently
getting from Eq. (\ref{rsquared}) the standard expression for the MB reflection probability
%
%%%%%%%%%%%%%%%%%%%%%%%%%%%%%%%%%%
\begin{equation}\label{rsquaredstandard}
|r|_{std}^2 \approx \frac{\pi |V_{12}|^2}{\sigma |v_{x 0}v_{y 0}|} \sim \frac{|V_{12}|^2}{\hbar \omega_c \varepsilon_F},
\end{equation}
%%%%%%%%%%%%%%%%%%%%%%%%%%%%%%%%%%
%
where $\textbf{v}_{0}$ is the electron velocity in the MB region. Comparing expressions (\ref{rsquared1}) and (\ref{rsquaredstandard}) shows that our configuration yields additional large parameter comparing to the standard situation, i.e. $|r|^2 = |r|_{std}^2 (\varepsilon_F / \hbar \omega_c)^{1/3}$, meaning that proportionally lower magnetic field is now required to achieve the same effect. \\

\section{Fourier series expansion of the density of states} \label{fourierapp}

The right-hand side of Eq.(\ref{density}) is a periodic function of
the phase $\Phi$ and hence it can be expanded in a Fourier series
%
%%%%%%%%%%%%%%%%%%%%%%%%%%%%%%%%%%
\begin{eqnarray}
\nu(E)=\frac{2 |S_+^\prime|}{(2\pi \hbar)^2}
\sum_{k=-\infty}^{+\infty}A_k e^{i k \Phi(E)},
\label{Fourier}
\end{eqnarray}
%%%%%%%%%%%%%%%%%%%%%%%%%%%%%%%%%%
%
where the Fourier coefficients are
%
%%%%%%%%%%%%%%%%%%%%%%%%%%%%%%%%%%
\begin{eqnarray}
A_k =  \int_{-\pi }^{\pi } \frac{e^{-ik\phi}|\sin
\phi|\Theta(|t|^2 -\cos^2\phi)}{\sqrt{|t|^2 -\cos^2\phi}} \frac{d
\phi}{2\pi }.
\label{Ak}
\end{eqnarray}
%%%%%%%%%%%%%%%%%%%%%%%%%%%%%%%%%%
%
Reducing the interval of integration to $(0, \pi)$, one finds
%
%%%%%%%%%%%%%%%%%%%%%%%%%%%%%%%%%%
\begin{eqnarray}
A_k&=& \left(1+(-1)^k  \right) \nonumber \\
&\times& \int_{0}^{\pi } \frac{e^{-ik\phi}|\sin \phi|\Theta(|t|^2
-\cos^2\phi)}{\sqrt{|t|^2 -\cos^2\phi}} \frac{d \phi}{2\pi },
\label{A2s}
\end{eqnarray}
%%%%%%%%%%%%%%%%%%%%%%%%%%%%%%%%%%
%
that is only even Fourier harmonics with $k=2 l, \;l=0,\pm 1,
\pm2,...$ are not equal to zero:   $A_{2 l}\neq 0, \; A_{2 l+1} =0$.
Using the equality
%
%%%%%%%%%%%%%%%%%%%%%%%%%%%%%%%%%%
\begin{eqnarray}
\cos2l\phi=\sum_{p=0}^l C_{2p}^l (-1)^{l-p}\cos^{2p}\phi
\sin^{2(l-p)}\phi,
 \label{cos}
\end{eqnarray}
%%%%%%%%%%%%%%%%%%%%%%%%%%%%%%%%%%
%
after rather simple transformation  one finds
%
%%%%%%%%%%%%%%%%%%%%%%%%%%%%%%%%%%
\begin{eqnarray}
A_{2l}= A_{-2l}=\frac{2}{\pi}\sum_{p=0}^l
C_{2p}^{2l}(-1)^{l-p}J_p^{(l)}, \;\;l=0,1,2,...
 \label{A2sfinal}
\end{eqnarray}
%%%%%%%%%%%%%%%%%%%%%%%%%%%%%%%%%%
%
where
%
%%%%%%%%%%%%%%%%%%%%%%%%%%%%%%%%%%
\begin{eqnarray}
J_p^{(l)}=|t|^{2p}\int_0^1\frac{x^{2p}(1-|t|^{2}x^2)^{l-p}}{\sqrt{1-x^2}}dx.
\label{J}
\end{eqnarray}
%%%%%%%%%%%%%%%%%%%%%%%%%%%%%%%%%%
%
In particular, from here and Eq.(\ref{A2sfinal}) it follows that
%
%%%%%%%%%%%%%%%%%%%%%%%%%%%%%%%%%%
\begin{eqnarray}
A_0 =1, \;A_{2}=|t|^2-1 =-|r|^2.
\label{A2}
\end{eqnarray}
%%%%%%%%%%%%%%%%%%%%%%%%%%%%%%%%%%

\end{document}